\documentstyle[11pt,aaspp4,amssym]{article}

\lefthead{Trager et al.}
\righthead{Galaxies at $z\approx 4$}

\begin{document}

\title{Galaxies at $z\approx4$ and the Formation of Population
II\altaffilmark{1}}

\author{S. C. Trager\altaffilmark{2}, S. M. Faber}
\affil{UCO/Lick Observatory and Board of Studies in Astronomy and
Astrophysics, University of California, Santa Cruz}
\authoraddr{Santa Cruz, CA 95064; sctrager{\@@}ociw.edu,
faber{\@@}ucolick.org}

\author{Alan Dressler \& Augustus Oemler, Jr.}
\affil{Observatories of the Carnegie Institution of
Washington}
\authoraddr{813 Santa Barbara Street, Pasadena, CA 91101;
dressler,oemler{\@@}ociw.edu}

\altaffiltext{1}{Based on observations taken with the NASA/ESA {\it
Hubble Space Telescope\/} obtained at the Space Telescope Science
Institute, which is operated by AURA under NASA contract NAS5-26555,
and observations obtained at the W. M. Keck Observatory, which is
operated jointly by the University of California and the California
Institute of Technology}

\altaffiltext{2}{Present address: Observatories of the Carnegie
Institution of Washington, 813 Santa Barbara Street, Pasadena, CA
91101}

\begin{abstract}
We report the discovery of four high-redshift objects ($3.3 < z < 4$)
observed behind the rich cluster CL0939+4713 (Abell 851).  One object
(DG 433) has a redshift of $z=3.3453$; the other three objects have
redshifts of $z\approx 4$: A0 at $z=3.9819$, DG 353 and P1/P2 at
$z=3.9822$.  It is possible that all four objects are being lensed in
some way by the cluster, DG 433 being weakly sheared, A0 being
strongly sheared, and DG 353 and P1/P2 being an image pair of a common
source object; detailed modelling of the cluster potential will be
necessary to confirm this hypothesis.  The weakness of common stellar
wind features like \ion{N}{5} and especially \ion{C}{4} in the spectra
of these objects argues for sub-solar metallicities, at least as low
as the SMC.  DG 353 and DG 433, which have ground-based colors, are
moderately dusty [$E_{\rm int}(B-V)\la 0.15$], similar to other $z>3$
galaxies.  Star formation rates range from $2.5\,(7.8)\,h^{-2}$ to
$22.\,(78.)\,h^{-2}\,M_{\odot}\ {\rm yr^{-1}}$, for $q_0=0.5\,(0.05)$,
depending on assumptions about gravitational lensing and extinction,
also typical of other $z>3$ galaxies.  These objects are tenatively
identified as the low-metallicity proto-spheroid clumps that will
merge to form the Population II components of today's spheroids.
\end{abstract}

\keywords{galaxies: formation --- galaxies: clusters: individual
(CL0939+4713) --- gravitational lensing}

\section{Introduction}

Identifying the precursors of today's old stellar
populations---ellipticals and bulges of spirals---is one of the major
goals of observational cosmology.  Until recently, we have been
confined to studying either galaxies at lookback times of less than
half a Hubble time or extremely distant, luminous objects such as
radio galaxies or QSOs. With the advent of 8--10m class telescopes,
faint star-forming galaxies at $2.5>z>4.5$ are now accessible.
Steidel et al.\ (1996a, 1996b; \cite{GSM96}) have pioneered a multicolor
technique to select candidates by use of color ``dropouts.''  At least
17 objects at $z>2.5$ have been found in the Hubble Deep Field using
this technique as of this writing (\cite{SGDA96}; \cite{Lowenthal96}).
Other groups have had success searching for QSO companions
(\cite{MTM96}; \cite{HM96}; \cite{HME96}; \cite{PPVC96}) or using
gravitational lensing to improve detectability (\cite{Ebbels96}).
Some high-redshift objects have been discovered serendipitously in
deep redshift surveys (\cite{Yee96}, \cite{EYBE96}).

In this paper we report observations of four high-redshift objects
($3.3<z<4$) observed through the intermediate-redshift cluster
CL0939+4713 (Abell 851, $z\approx 0.4$).  These objects were
discovered serendipitously during a program to study the stellar
populations of early-type galaxies in this cluster.  CL0939+4713 has
been the subject of intensive study over the past 15 years
(\cite{DG92}; Dressler et al.\ 1993, 1994a,b; \cite{Moore96};
\cite{Trager97}) as an excellent laboratory for studying the
evolutionary state of galaxies at $z=0.41$.  It has also been the
subject of a weak-field gravitational lensing study by Seitz et al.\
(1996), who identified three of our four objects as candidate lensed
galaxies.

We use the properties of our newly discovered objects, together with
those of other high-redshift samples, to argue that high-redshift
galaxies are metal-poor subclumps that will merge to form the
Population II components of today's spheroids.

\section{Observations}

\begin{figure*}[p]
\plotfiddle{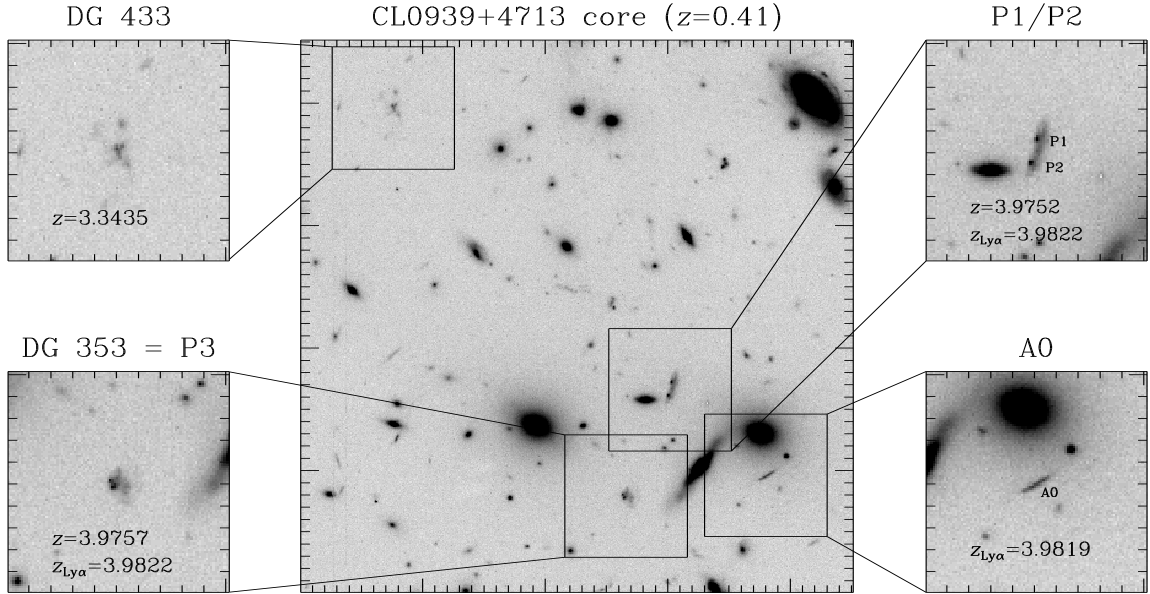}{3truein}{-90}{75}{75}{-262}{416}
\caption[The core of CL0939+4713 as imaged by the {\it Hubble Space
Telescope}]{The core of CL0939+4713 as imaged by the {\it Hubble
Space Telescope.\/} The bright galaxies are cluster members at $z=0.41$;
the fainter galaxies are a combination of faint dwarf galaxies in the
cluster and very distant galaxies behind the cluster.  The central
panel is 140 $h^{-1}$ kpc on a side at the distance of the cluster.
The four side panels are expanded views of four very distant galaxies
behind the cluster.  Each panel is about 30 $h^{-1}$ kpc across in the
rest frame of the objects (assuming no lensing), and the resolution is
about 0.3 $h^{-1}$ kpc.  North is to the top, east is to the left; the
small ticks are $1\arcsec$ apart.} \label{figure1}
\end{figure*}

The observations were taken as part of a multislit spectral survey of
objects in CL0939+4713 and the superposed field.  Two faint,
morphologically peculiar objects (DG 353, DG 433; \cite{DG92}) were
selected near the core of the cluster from the HST images of Dressler
et al.\ (1994b), Figure~\ref{figure1}, to sample what we then believed
to be dwarf irregular cluster members.  These and about twenty other
objects were observed through a multi-aperture plate with the LRIS
spectrograph (\cite{Oke95}) for 7200 s on 18 March 1995.  The plates
were punched with slit widths of $0.7$ arcsecond.  The seeing was
$0\farcs7$ FWHM, the instrumental resolution was $\approx 7.5$ \AA,
and the spectral range was 3700--9000 \AA.

While preparing for a second Keck run in March 1996, we reduced and
re-examined these spectra in early February 1996 using the EXPECTOR
spectral extraction package (\cite{Kelson97}).  The spectra of DG 353
and DG 433 proved not to be those of dwarf galaxies at $z\approx 0.4$;
rather, they resembled the spectra of high-redshift ($z>3$)
star-forming galaxies discovered by Steidel et al.\ (1996a).

Based on these results, we selected several more faint,
morphologically peculiar objects. These included two arc-like objects
recently suggested as possibly gravitationally lensed and distorted
images of distant background galaxies (\cite{SKSS96}; these are P1/P2
and A0 in their notation, see Figure~\ref{figure1}; in their notation,
DG 353 is also known as P3).  We used tilted slitlets on the
multi-aperture plates to cover multiple objects (in the case of P1/P2)
or to follow the major axis of the object (A0).  The plate with a
slitlet for P1/P2 was exposed for a total of 6382 s on 16 March 1996,
and the plate with a slitlet for A0 was exposed for a total of 7200 s
on 16--17 March 1996.  The instrumental configuration was identical to
the March 1995 run.  The seeing was slightly worse and variable,
ranging from $0\farcs7$ to $1\farcs1$.  Reductions were again
performed using the EXPECTOR spectral extraction package
(\cite{Kelson97}).

\begin{figure*}
\plotone{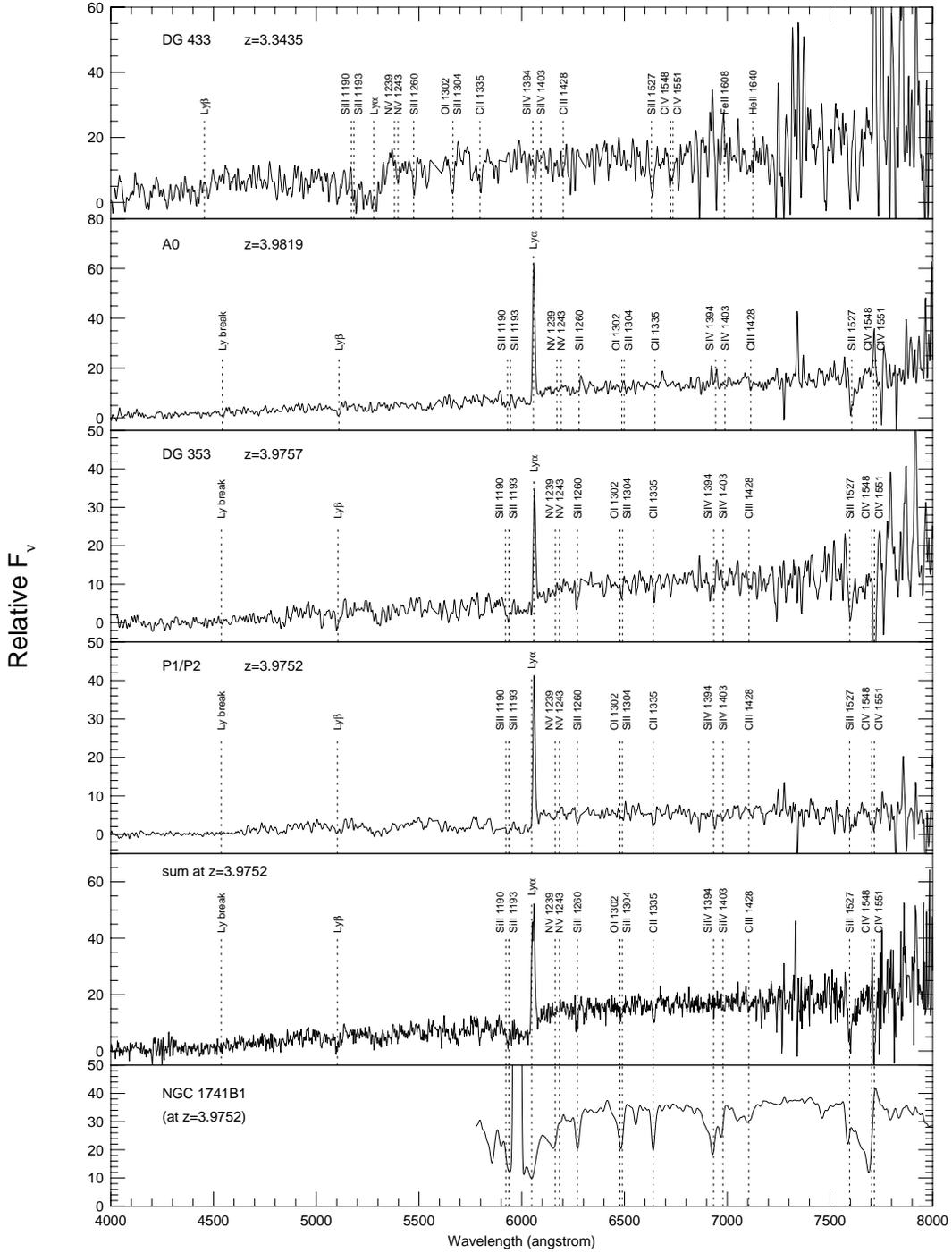}
\caption[Spectra of $z>3$ objects]{Spectra of the four $z>3$ objects.
The fifth panel is a sum of the four objects, corrected to a common
redshift.  A spectrum of NGC 1741B1 (kindly provided by C. Leitherer),
a bright knot in a local starburst, has been redshifted and smoothed
to match the object spectra.} \label{figure2}
\end{figure*}

Spectra of the four objects are shown in Figure~\ref{figure2}.  A sum
of the four objects (below) is shown in the fifth panel, and a
spectrum of NGC 1741B1, a knot in a nearby starburst galaxy
(\cite{CLV96}), kindly provided by C. Leitherer, is shown in the sixth
panel for comparison, redshifted and convolved to the resolution of
the object spectra.

\section{Results}

\subsection{Redshifts}

Redshifts were measured independently for these objects from detected
absorption lines and from Ly$\alpha$ emission (not detected in DG 433).
These redshifts are summarized in Table 1.

\begin{deluxetable}{lccllccc}
\scriptsize
\tablecaption{Object properties.\label{table1}}
\tablewidth{0pt}
\tablehead{
\colhead{} & \colhead{} & \colhead{} & \colhead{} & \colhead{} &
\colhead{peak SB\tablenotemark{d}} & \colhead{} &\colhead{} \\
\colhead{Name} & \colhead{$z_{abs}$} & \colhead{$z_{\rm
Ly\alpha}$\tablenotemark{a}} & \colhead{$R_{702}^{\rm
iso}$)\tablenotemark{b}} & \colhead{$R_{702}^{\rm
Kron}$)\tablenotemark{c}} & \colhead{(mag/$\Box\arcsec$)} &
\colhead{area ($\Box\arcsec$)\tablenotemark{e}} & \colhead{axial ratio}
}
\tablecolumns{8}
\startdata
DG 433&3.3435&\nodata&24.46&23.31&22.28&0.25&0.69\nl
DG 353&3.9757&3.9822\tablenotemark{f}&23.27&22.78&21.76&0.59&0.84\nl
\phantom{DG 353}a& & & & &21.76& &\nl
\phantom{DG 353}b& & & & &21.85& &\nl
P1/P2&3.9752&3.9822&22.62&22.00&21.39&1.10&0.40\nl
\phantom{P1/}P1& & & & &21.39& &\nl
\phantom{P1/}P2& & & & &21.57& &\nl
A0&\nodata&3.9819\tablenotemark{f}&23.58&\nodata\tablenotemark{g}&22.13&0.49&0.21\tablenotemark{h}\nl
 & & & & & & &\nl
\enddata
\tablenotetext{a}{See Table 2}
\tablenotetext{b}{Isophotal magnitude at 23.5 mag/$\Box\arcsec$}
\tablenotetext{c}{Elliptical aperture magnitude with semimajor axis as
defined by Bertin \& Arnouts (1996), following Kron (1980); an
approximation to the total magnitude}
\tablenotetext{d}{Surface brightness of peak pixel}
\tablenotetext{e}{Isophotal area at 23.5 mag/$\Box\arcsec$}
\tablenotetext{f}{After correction for a zero-point shift and
wavelength scale stretching with respect to P1/P2; see Section 3.2}
\tablenotetext{g}{Contaminated by neighboring large elliptical DG 367
(Dressler \& Gunn 1992)}
\tablenotetext{h}{A0 only marginally resolved along short axis}
\end{deluxetable}

We detect weak low-ionization interstellar absorption lines of
\ion{Si}{2}, \ion{C}{2}, and possibly \ion{O}{1} in three of these
objects (DG 433, DG 353, and P1/P2).  Higher-ionization lines of
\ion{C}{4} and \ion{Si}{4} may also be detected in some of the
objects, but they are even weaker.  At these levels, all or most of
their equivalent widths may also be interstellar; evidence for truly
photospheric stellar lines is tenuous.  From detected {\it
absorption\/} lines, DG 353 has a redshift of $z=3.9757\pm 0.0026$ (7
lines), DG 433 has a redshift of $z=3.3435\pm 0.0026$ (6 lines), and
P1/P2 has a redshift of $z=3.9752 \pm 0.0039$ (7 lines).  The
absorption-line redshift of P1/P2 is consistent with that of DG 353.
The quoted errors are the RMS errors of the means based on the scatter
in all detected absorption lines.  Only the \ion{Si}{2}--\ion{C}{4}
absoprtion complex is potentially detected in A0; its redshift is
consistent with the Ly$\alpha$ redshift, though the region is heavily
contaminated by poor subtraction of night-sky emission.

Ly$\alpha$ in emission is detected in all objects except DG 433.  From
Ly$\alpha$ alone, the formal redshift of DG 353 is $z=3.9837 \pm
0.0001$, that of P1/P2 is $z=3.9822 \pm 0.0001$, and that of A0 is
$z=3.9812 \pm 0.0001$ (errors are centering errors only).  Note that
for DG 353, this Ly$\alpha$ redshift implies a velocity difference
with respect to the absorption lines of $480\ {\rm km\,s^{-1}}$ in the
rest frame of the object; for P1/P2, the implied velocity difference
is $420\ {\rm km\ s^{-1}}$.  The formal velocity difference between
DG 353 and P1/P2 is $\Delta z=0.0015\pm0.0001$, about $80\ {\rm km\
s^{-1}}$ in the restframe.

\begin{deluxetable}{llrrrr}
\small
\tablecaption{Cross-correlation velocity differences.\label{table2}}
\tablewidth{0pt}
\tablehead{
\colhead{} & \colhead{} & \colhead{$\Delta v$ ($\rm km\
s^{-1}$)} & \colhead{$\Delta v$ ($\rm km\ s^{-1}$)} & \colhead{$\Delta
v$ ($\rm km\ s^{-1}$)} & \colhead{$\Delta v_{corr}$ ($\rm km\
s^{-1}$)} \\ 
\colhead{Name} & \colhead{Name} &\colhead{Na D $\lambda5893$} & \colhead{O I
$\lambda6300$} & \colhead{Ly$\alpha$} &
\colhead{Ly$\alpha$\tablenotemark{a}}}
\tablecolumns{6}
\startdata
DG 353&P1/P2&$75\pm4$&$101\pm13$&$80\pm44$&$6\pm45$\nl
A0&P1/P2&$32\pm2$&$40\pm5$&$119\pm44$&$83\pm44$\nl
\enddata
\tablenotetext{a}{After correction for wavelength zero-point differences and
stretchings.}
\end{deluxetable}

The $15\sigma$ formal difference between the Ly$\alpha$ redshifts of
DG 353 and P1/P2, which have consistent absorption redshifts, appears
to be an artifact of the wavelength calibration process.  In order to
check our wavelength solutions between exposures and runs, we
performed cross-correlations of the wavelength-calibrated, rectified
spectra before sky-subtraction. Using the RV.FXCOR task in IRAF, we
cross-correlated the spectra of DG 353 and P1/P2 in 200\AA\ regions
around both the Na~D $\lambda 5893$ and \ion{O}{1} $\lambda 6300$
night-sky emission lines.  The measured velocity differences are
$75\pm4\ {\rm km\ s^{-1}}$ and $101\pm13\ {\rm km\ s^{-1}}$,
respectively, indicating a slight wavelength stretch.
Cross-correlation of a 100\AA\ region around Ly$\alpha$ at $\approx
6060$\AA\ found a velocity difference of $80\pm6\ {\rm km\
s^{-1}}$. This error is the formal error from the cross-correlation
routine; however, the line profiles are skewed, and we estimate the
true uncertainty to be closer to $\pm44\ {\rm km\ s^{-1}}$, or
one-tenth of the FWHM of Ly$\alpha$ in these objects. After correction
for both the zero-point difference and wavelength stretching, the
velocity difference between the Ly$\alpha$ lines of DG 353 and P1/P2
is only $6\pm45\ {\rm km\ s^{-1}}$.  The $80\ {\rm km\ s^{-1}}$
correction is small compared to the errors of the absorption-line
redshifts and does not disturb their previous agreement.  The
absorption-line redshifts are derived from lines over a broad spectral
region, and local wavelength shifts are not important, in contrast to
Ly$\alpha$. P1/P2 and DG 353 thus have consistent absorption-line and
emission-line redshifts.  Their spectra are also consistent.

We performed a similar cross-correlation analysis on the spectra of
P1/P2 and A0 (taken on the same run).  The measured velocity
differences for Na~D $\lambda 5893$ and O~I $\lambda 6300$ are
$32\pm2\ {\rm km\ s^{-1}}$ and $40\pm5\ {\rm km\ s^{-1}}$,
respectively.  Cross-correlation of the Ly$\alpha$ regions determined
a velocity difference of $119\pm3\ {\rm km\ s^{-1}}$, but the
uncertainty is again probably closer to $\pm44 {\rm km\
s^{-1}}$. After correction for the difference in wavelength
calibrations, the velocity difference between the Ly$\alpha$ lines of
A0 and P1/P2 is $83\pm44\ {\rm km\ s^{-1}}$, indicating a
marginally significant difference.  The spectra of these objects are
also different; P1/P2 clearly has interstellar absorption lines, while
A0 appears to be line-free.  Table 2 presents a summary of the
velocity differences as determined by cross-correlation.

\subsection{Photometry}

Isophotal magnitudes and areas were measured from stacked,
geo\-metrical\-ly-cor\-rec\-ted WFPC2 F702W (hereafter called
$R_{702}$) images of Dressler et al.\ (1994b) using the SExtractor
faint-galaxy photometry package (\cite{BA96}).  Figure~\ref{figure1}
shows the core of the cluster, in WF2.  Magnitudes were calibrated
using the ``synthetic'' WFPC2 system of Holtzman et al.\ (1995).  The
$3\sigma$ surface brightness limit in this frame from Poisson noise of
the sky is $\approx 25.9\ \rm mag/\Box\arcsec$; however, a relatively
high surface brightness threshold of $23.5\ \rm mag/\Box\arcsec$ was
applied, as A0 is blended with the large nearby elliptical DG 367
(\cite{DG92}) at fainter levels.  Elliptical aperture magnitudes
approximating total aperture magnitudes were also measured following a
modified version of the Kron (1980) $r_1$ aperture magnitude scheme,
as described by Berin \& Arnouts (1996). Table 1 summarizes the
properties of these four objects, including redshifts, $R_{702}$
isophotal and Kron aperture magnitudes, peak surface brightnesses,
isophotal areas, and axial ratios.

\section{Gravitational lensing}

Seitz et al.\ (1996) have suggested that P1/P2 and DG 353 are a lensed
pair and that A0 is a gravitational arc.  The coincident redshifts of
P1/P2 and DG 353 and their similar spectra and double-knot
morphologies support this suggestion, but more detailed modelling of
the cluster potential will be necessary to determine if these objects
do indeed share a common source.  A0, with its slightly discrepant
redshift, line-free spectrum, and smooth morphology, may also be
lensed but probably is not another image of the possible source galaxy
of P1/P2/DG 353.

DG 433 is far enough outside the cluster core,
$35\arcsec=161\,(176)\,h^{-1}\;\rm kpc$ [$q_0=0.5\,(0.05)$] projected
distance from the three dominant elliptical galaxies, that it is not
expected to be greatly distorted or amplified by the cluster
potential.  The mean ellipticity expected at the position of DG 433
from Figure 3 of Seitz et al.\ (1996) is
$\|\overline{\epsilon}\|\approx0.083$.  This corresponds to a
stretching of about 18\% (roughly along the major axis) and a
brightening of about $0.18$ mag.  The HST image of this galaxy is thus
more likely to be a truer representation of its ultraviolet morphology
than those of the other three objects.

\section{Object properties}

\subsection{Metallicities}

As pointed out by Pettini \& Lipman (1995) and Steidel et al.\
(1996a), it is difficult to determine precisely the metallicities of
galaxies using low-resolution spectra of far-UV absorption lines.
However, one can make a very crude estimate by comparing to the
stellar absorption lines in nearby O stars and starburst galaxies.
Walborn et al.\ (1995) note that the high-excitation lines \ion{N}{5},
\ion{Si}{4}, and \ion{C}{4} are produced mostly in stellar
photospheres, and their strength is an indicator of stellar surface
abundance.  This is confirmed by comparing UV spectra of O stars from
the SMC ([Fe/H]$\;=-0.65$, \cite{RB89}) with similar stars from the
LMC ([Fe/H]$\;=-0.30$).  \ion{N}{5} is occasionally strong in SMC
stars, probably due to dredge-up of CNO processed material from the
interior (\cite{MC94}), but all three lines would certainly be weak in a
composite UV spectrum of the SMC.  These features are also uniformly
quite strong in the solar-abundance starburst models of Leitherer,
Robert, \& Heckman (1995).

The fifth panel of Figure~\ref{figure2} shows a composite sum of the
four distant galaxy spectra presented here.  \ion{N}{5}, \ion{Si}{4},
and \ion{C}{4} are all weak or absent.  The weakness of \ion{C}{4} is
particularly telling.  In local O stars and starburst galaxies, it is
the most constant and reliable stellar line, and the starburst
spectrum of NGC 1741B1 (bottom panel of Figure~\ref{figure2};
[O/H]$\;=-0.61$, \cite{VC92}) is an indicator of the expected strength
of \ion{C}{4} near solar metallicity.  \ion{C}{4} is contaminated by
strong sky lines in three of our four objects, but nevertheless it is
clear that its strengh in the sum is much weaker than in NGC 1741B1.
Finally, the near-total lack of {\it any\/} line in A0 (including
interstellar lines, which are strong in the SMC despite its low
metallicity) suggests an {\it extremely\/} low metallicity for this
object.

The average stellar line strength of our objects is strikingly
similar, or perhaps even weaker, than that of the composite spectrum
of 11 high-redshift galaxies shown by Lowenthal et al.\ (1997).
Futhermore, including the spectra of Steidel and collaborators
(\cite{SGPDA96}, 1996b), a total of more than 25 galaxies at $z>2.5$
have been observed so far, yet {\it not one\/} has shown stellar line
strengths approaching those of the local solar-metallicity starburst
NGC 1741.  An exact estimate of the metallicities of high-redshift
galaxies awaits higher signal-to-noise spectra as well as
non-solar-abundance starburst models (Robert et al., in preparation).
However, a rough comparison suggests that the average mean metallicity
of these objects is at least as low as the SMC, {\it i.e.,} $Z\la
0.1Z_{\odot}$.  Thus, we suggest that high-redshift galaxies as a
class are likely to be metal poor.

\subsection{Dust}
\label{sec:dust}

A rough estimate of the reddening of DG 433 and DG 353 can be made by
appealing to their ground-based $g-r$ colors: $1.09$ and $1.24$,
respectively (Gunn \& Dressler, unpublished; the ground-based colors
of P1/P2 and A0 are contaminated by nearby objects).  Using unreddened
models of very young star-forming galaxies observed at the appropriate
redshifts (\cite{BC93}; continous star-forming galaxy models of 10
and 50 Myr kindly provided by C. Gronwall) and taking into account the
blanketing of the $g$ band by the Ly$\alpha$ forest calculated by
Madau (1995), one expects $g-r\approx 0.1$ and $\approx 0.8$ for DG
433 and DG 353, respectively.  These colors imply reddenings of
$E(g-r)\approx 1.0$ and $\approx 0.4$ mag.  For a simplified
foreground screen model and an SMC extinction law (\cite{CKS94})
appropriate to low metallicities, these values correspond to
extinctions at observed $r$ (rest 1500 \AA\ and 1300 \AA,
respectively) of $1.9$ and $0.9$ magnitudes.  This corresponds to
optical restframe internal reddenings of $E(B-V)=0.15$ and $0.06$
mag. These estimates are rough owing to the use of a foreground screen
model and uncertainties in the reddening curve.  However, they suggest
that the observed UV luminosities may need to be corrected upwards by
modest factors (2--6).  This amount of UV absorption is adequate to
reduce Ly$\alpha$ to the modest levels generally seen in high-redshift
galaxies (\cite{CF93}).

\subsection{Star Formation Rates}

In the absence of large dust extinction, star formation rates can be
estimated directly from the rest-frame far-UV flux observed in the
{\it Hubble Space Telescope\/} images.  A $\sim 7$ Myr old
``continuous star formation'' model from the evolutionary-synthesis
models of Leitherer, Robert, \& Heckman (1995) with $Z=Z_{\odot}$
produces a rest-frame luminosity at 1500 \AA\ of $L_{1500}=10^{40.1}\
{\rm erg\ s^{-1}\ \AA^{-1}}$ (models with $Z=0.1 Z_{\odot}$ have
similar continuum luminosities; Robert, private communication).  The
effective wavelengths of the F702W filter at $z=3.35$ and $z=3.98$ are
$\approx 1600$\ \AA\ and $\approx 1400$\ \AA; to compute $L_{1500}$,
we note that, redwards of Ly$\alpha$, $f_{\lambda}\propto\lambda^{-2}$
for DG 433, P1/P2, and DG 353, and $f_{\lambda}\propto\lambda^{-2.5}$
for A0.  The isophotal F702W magnitudes $R_{702}^{\rm iso}$ (Column 4
of Table~\ref{table2}) imply raw star formation rates, ignoring
possible gravitational lensing, of $2.5\,(7.8) \,h^{-2}\,M_{\odot}\
{\rm yr^{-1}}$ (DG 433), $7.1\,(25.)  \,h^{-2}\,M_{\odot}\
{\rm yr^{-1}}$ (A0), $18.\,(63.)  \,h^{-2}\,M_{\odot}\
{\rm yr^{-1}}$ (P1/P2), and $9.8\,(31.)  \,h^{-2}\,M_{\odot}\
{\rm yr^{-1}}$ (DG 353) for $q_0=0.5$ ($q_0=0.05$).  Correcting for
the possible weak-field lensing of DG 433 mentioned above would reduce
its star formation rate to $2.1\,(6.6) \,h^{-2}\,M_{\odot}\
{\rm yr^{-1}}$, and the star formation rates of A0, P1/P2 and DG
353 are upper limits in the absence of a lensing correction.
Accounting for extinction in DG 433 would raise its SFR to $12.\,(38.)
\,h^{-2}\,M_{\odot}\ {\rm yr^{-1}}$, including the weak-field
lensing; for DG 353, accounting for extinction would raise its SFR to
$22.\,(78.) \,h^{-2}\,M_{\odot}\ {\rm yr^{-1}}$, ignoring any
lensing.

To summarize, an average {\it raw\/} star formation rate of
$9.3\,(32.)  \,h^{-2}\,M_{\odot}\ {\rm yr^{-1}}$, plus an average
correction factor of $\sim 4$ for extinction, yields a corrected
average rate of $\approx 32.\,(120.) \,h^{-2}\,M_{\odot}\
{\rm yr^{-1}}$.  The raw rate is higher than the average raw star
formation rate measured by Lowenthal et al. (1997),
$\dot{M}=1.7\,(4.7) \,h^{-2}\,M_{\odot}\ {\rm yr^{-1}}$ based on 11
galaxies.  However, three of the four objects here may be
significantly lensed, and their inferred star formation rates are
therefore upper limits.

Sizes of these objects are difficult to measure due to possible
lensing, but the raw half-light radii are $\lesssim 0\farcs5$,
corresponding to $\lesssim 8. (15.) h^{-1}$ kpc for $q_0=0.5
(0.05)$. These are comparable to the radii seen among other
high-redshift objects (\cite{GSM96}; \cite{Lowenthal96}).

\section{Implications}

From a handful of early studies, a portrait of high-redshift
``normal'' galaxies is beginning to emerge (\cite{SGPDA96}, 1996b;
\cite{GSM96}; \cite{Lowenthal96}).  They have star-formation rates of
1--10 $h^{-2}\,M_{\odot}\ {\rm yr^{-1}}$, rest frame $B$-band
luminosities of 1--few $L^{\ast}$, and half-light radii of 1--10 kpc.
Morphologically, they usually consist of compact,
high-sur\-face-bright\-ness blobs, typically single, sometimes
multiple, and sometimes imbedded in lower-sur\-face-bright\-ness,
diffuse luminosity.  The co-moving number density of high-redshift
galaxies now approaches or exceeds the local number density of
$L>L^{\ast}$ galaxies (\cite{Lowenthal96}).

What kind of local objects will these high-redshift galaxies evolve
into, and what kind of snapshot of early galaxy formation do they
provide?  Steidel et al.\ (1996a, 1996b) and Giavalisco et al.\ (1996)
have suggested that they may represent the {\it cores\/} of future
massive spheroids.  They cite the smooth, compact morphologies of
their sample, plus a tentative estimate of virial motions $\sim 200\
{\rm km \ s^{-1}}$ based on the equivalent widths of interstellar
absorption lines.  As the lookback time to $z=3.5$ is 13.5 (12.5) Gyr
for $t_0=15$ Gyr and $q_0=0.5$ (0.05), the stars in high-redshift
galaxies are plausibly identified with local old spheroid populations.
Finally, the presence of compact, sometimes multiple blobs and diffuse
emission indicates that the stars that have already formed will not
ultimately reside in flattened, rotating disks.  Either the blobs will
remain isolated or they will merge with other blobs.  In either, case
the stars that we now see will populate a spheroidal density
distribution, strengthening the identification with local spheroids.

Despite these parallels, further data suggest that the massive core
hypothesis may need some revision.  First, the equivalent widths of
the interstellar lines have been challeged as indicators of internal,
gravitational motions (\cite{CLV96})---saturation effects are just too
uncertain.  It seems fair to say that the internal motions of
high-redshift galaxies are presently unknown---these galaxies could be
massive or not.  Second, strong redshift clustering has not been
detected for the high-redshift population (\cite{Lowenthal96};
\cite{SGPDA96}), except possibly in the fields of QSOs (\cite{MTM96};
\cite{HME96}; \cite{HM96}).  The high-redshift population is thus
likely to be a {\it field\/} population, not the precursor of
clusters.  This, together with the large number density approaching or
exceeding the local value of $N(L>L^{\ast})$, suggests that these
objects are not the precursors of {\it ellipticals\/} but rather the
spheroids of early-type (mostly field) {\it spirals,} which are
locally much more numerous than ellipticals.

Equating high-redshift galaxies with local spheroids rather than
massive ellipticals goes some way towards solving a size discrepancy
that results from the latter hypothesis.  A massive elliptical like
M87 would have an apparent half-light radius $\approx 0\farcs9\,h$
(for $q_0=0.5$ and an age of 1 Gyr) at $z=3$.  This is five to ten
times larger than the typical half-light radii of the compact
high-redshift blobs, which are more consistent with those of local
bulges (\cite{BBF92}, as noted by \cite{Lowenthal96}).

However, this seeming agreement conceals a serious problem, as
Lowenthal et al.\ (1997) note.  The observation that the stars of
high-redshift galaxies are {\it metal-poor\/} suggests that we should
properly be comparing high-redshift radii to the radii of the
metal-poor ({\it i.e.,} Population II) components of local spheroids.
These are larger than the half-light radii given above---in the Milky
Way, for example, the halo globular cluster population with
$Z<0.1Z_{\odot}$ has a mean radius of 10 kpc (\cite{HR79}) and
extends to $\sim 50$ kpc.  These are also much larger than the typical
spheroid half-light radius $r_{1/2}\sim2$ kpc (\cite{BBF92}).  The
half-light radius of the Population II field stars in the Milky Way is
not known, but since the bulge of the Galaxy is metal-rich
(\cite{MR94}; \cite{SRT96}), there can be no doubt that $r_e$ for
the metal-poor stars is much larger than $r_e$ for the bulge.  The
identification of high-redshift stars with local metal-poor Population
II components tends to reintroduce the size discrepancy.

We suggest that the resolution of this discrepancy may be found in the
currently popular theory for the formation of spheroids via
hierachical clustering and merging.  Hierarchical clustering predicts
that highly overdense systems like those believed to form spheroids
and bulges begin to collapse with many subclumps (\cite{BFPR84}).
Searle \& Zinn (1978) presented a subclump formation model of the
metal-poor spheroid of the Milky Way that in many respects is similar.
The essence of these pictures is that the first (metal-poor) stars are
formed in small clumps distributed {\it throughout\/} the volume that
will later become the Population II spheroid.  As these clumps merge,
they create a large, diffuse halo of metal-poor stars.  Meanwhile,
enriched gas continues to dissipate and fall to the center of the
global potential well, giving rise to successive generations of more
metal-rich and more centrally concentrated stars.  This is essentially
a clumpy version of the collapse picture for the Milky Way spheroid
advocated by Eggen, Lynden-Bell, \& Sandage (1962) and seen now in
more realistic hydrodynamic simulations ({\it e.g.,} \cite{SM95}).
Thus, spheroids do not form from the inside out, as massive cores
accreting matter to well-defined centers, but from the outside in,
with the first stars more widely distributed than the later ones.  We
suggest that the compact blobs of the high-redshift galaxies should be
equated with the subclumps of the Searle-Zinn and hierarchical
clustering models.  In that case, their small radii pose no
contradiction.  This interpretation was first put forward as one of
several scenarios by Lowenthal et al.\ (1997).

The high co-moving number density of high-redshift galaxies is
comparable to the local co-moving number density of spheroids
(\cite{Lowenthal96}).  If the above identification with local
spheroids is correct, this implies that we are seeing roughly {\it
all\/} forming spheroids at high-redshift.  In other words, the
structure in and around the luminous blobs must be a statistically
representative snapshot of what an early spheroid looked like.  Given
the present size of the Population II component in the Milky Way, we
might therefore expect to see a {\it population\/} of subclumps in
every high-redshift spheroid, distributed over a volume $\gtrsim 20$
kpc in diameter, or $\gtrsim 1\farcs4$ ($0\farcs8$) on the sky for
$q_0=0.5$ (0.05).  We do not see that.  Instead, we see most often
only one blob, or at most a few blobs, in a significantly smaller
volume.  There are a handful of cases (\cite{Lowenthal96};
\cite{GSM96}) where companion blobs separated by $\sim1\arcsec$ have
colors that suggest similar redshifts, but it seems possible that
there are few {\it large populations\/} of bright blobs filling such
volumes.

This contradiction might be resolved if each individual blob was
bright for a short time, so that, on average, {\it only one blob is
seen at a time.\/} We would then have a ``Christmas tree'' model
(\cite{Lowenthal96}), wherein small individual star-forming blobs come
and go within a much larger and largely invisible spheroid structure.
Semi-analytic models of galaxy formation (Somerville \& Primack,
priv.\ comm.) currently do not make blobs nearly as bright as the
$z>3$ galaxies seen.  However, these authors have suggested that
allowing blobs to collide with each other, triggering star formation,
might produce blobs occasionally bright enough to reproduce the $z>3$
galaxies observed.

Alternatively, dust could hide blobs.  Every
spectro\-scop\-ical\-ly-con\-firm\-ed $z>3$ galaxy to date has very
weak stellar absorption-line features suggesting low metallicity.  If
dust content were to increase with metallicity, clumps might
{\it self-obscure\/} as star-formation and nucleosynthesis proceeded
(\cite{CF93}).  If this process were rapid, only a few blobs might be
seen at a time.

We may now enquire whether either of these models is consistent with
known constraints on the timescale of spheroid formation.  Let $n_v$
be the number of blobs visible at any given time, and let $t_v$ be the
time a blob is visible.  Let $N_b$ be the total number of blobs needed
to form a spheroid of mass $M_S$ over total time $T_S$.  Then
\begin{equation}
n_v = \frac{t_v N_b}{T_S}.
\end{equation}
If $t_{\ast}$ is the star-formation timescale of a blob, the quantity
\begin{equation}
f=\frac{t_v}{t_{\ast}} \leq 1
\end{equation}
is the fraction of time that a blob is visible while forming stars.
If $M_b$ is a typical blob stellar mass and $R$ is its star-formation
rate, then
\begin{equation}
N_b M_b = M_S,\ M_b = R t_{\ast}
\end{equation}
and the number of blobs visible at any one time is
\begin{equation}
n_b = f\frac{M_S}{R T_S}.
\end{equation}
Thus, if $n_b\approx 1$ as observed and $f=1$, the timescale for
forming a $10^{10} M_{\odot}$ spheroid is
\begin{equation}
T_s \approx \begin{array}{r}
		3 \times 10^9\ {\rm yr} \\
		3 \times 10^8\ {\rm yr}
	 \end{array}
\ {\rm for}\ R = \begin{array}{r}
		 3\ M_{\odot}\ {\rm yr^{-1}} \\
		 30\ M_{\odot}\ {\rm yr^{-1}}
		 \end{array}
.
\end{equation}
For $f=0.1$, the timescale for forming a $10^{10} M_{\odot}$ spheroid
is
\begin{equation}
T_s \approx \begin{array}{r}
		3 \times 10^8\ {\rm yr} \\
		3 \times 10^7\ {\rm yr}
	 \end{array}
\ {\rm for}\ R = \begin{array}{r}
		 3\ M_{\odot}\ {\rm yr^{-1}} \\
		 30\ M_{\odot}\ {\rm yr^{-1}}
		 \end{array}
.
\end{equation}
The observed star formation rate for the blobs is $R\sim10\,M_{\odot}\
{\rm yr^{-1}}$, and an estimated $T_S$ for the Milky Way is
$\gtrsim2$ Gyr (\cite{SNPBO96}).  These seem more consistent with $f=1$
than $f=0.1$.  Thus, blobs may not self-obscure due to dust, at least
not at very early epochs.

Constraints on blob lifetimes are loose.  If blobs make stars only in
short bursts, $t_{\ast}$ could be as short as the minimum starburst
age of $\lesssim10$ Myr, the typical blob stellar mass $M_b$ would be
only $\sim10^8\, M_{\odot}$ for $R=10\,M_{\odot}\ {\rm yr^{-1}}$,
and there would be 100 blobs per $10^{10} M_{\odot}$ spheroid.  If
$t_{\ast}=10^9$ yr (a plausible upper limit), then the blob mass is
$10^{10}\,M_{\odot}$ and there is only one blob per spheroid.  This
would incompatible with the need for collisions among blobs to create
an extended Population II halo (Somerville \& Primack, priv.\ comm.).
Thus $t_{\ast}$ is most likely in the range $10^7$--$10^8$ yr.  The
difference corresponds to a factor of $\sim10$ in $M/L$ ratio
(\cite{BC93}) and might be detectible if linewidths (and masses) could
be measured.  A longer $t_{\ast}$ might also give rise to an older
stellar population detectible in the $K$-band.

\section{Summary}

We have presented spectra of four objects at $z>3$, three of which are
at $z\approx 4$.  These galaxies are
\begin{itemize}
\item small:  $r_{1/2}\approx 10$ kpc;
\item bright:  $L_B \sim 1$--$10\,L_B^{\ast}$;
\item lumpy, with no more than a few blobs per object;
\item likely metal-poor, with $Z\lesssim 0.1Z_{\odot}$;
\item mildly dusty:  $E(B-V)_{{\rm int}}\lesssim 0.15$;
\item and star forming: $\dot{M} \sim 2$--$60\, M_{\odot}\
{\rm yr^{-1}}$.
\end{itemize}
These particular galaxies probably have been magnified and brightened
somewhat by gravitational lensing but otherwise seem to be the
higher-redshift analogs of $z\sim3$ galaxies seen previously
(\cite{SGPDA96}, 1996b; \cite{Lowenthal96}).  Considering their
properties together with the properties of other galaxies at $z>3$,
particularly their metallicities, number density, star-formation
rates, and sizes, we suggest that these objects will evolve into the
analogs of spheroids, in particular the {\it Population II components
of early-type spirals\/}.  The individual blobs may be identified with
the Searle \& Zinn (1978) subclumps thought to have formed the Milky
Way halo.  We present a toy model of the formation of spheroids, with
particular emphasis on the number of star-forming blobs visible in a
given halo at any one time.  Kinematic data on blob masses could be
very helpful in testing this ``Christmas tree'' model.

\acknowledgments

We thank C. Grillmair and the CARA staff, particularly T. Bida,
J. Aycock, and W. Wack, for observing assistance, D. Kelson and
A. Phillips for software and support, C. Leitherer for providing the
spectrum of NGC 1741B1, C. Gronwall and P. Madau for providing digital
versions of model data, C. Robert for providing models in advance of
publication, I. Smail, J.-P.\ Kneib, C. Steidel, M. Giavalisco,
J. Lowenthal, E. Groth, and E. Shaya for helpful discussions, and an
anonymous referee for comments that substantially improved the
manuscript.  This work was partially supported by a Flintridge
Fellowship to SCT and by NASA contract NAS5-1661 to the WF/PC-I IDT.


\clearpage

\begin{thebibliography}{}
\bibitem[Bender, Burstein, \& Faber 1992]{BBF92} Bender, R., Burstein,
D., \& Faber, S. M.  1992, \apj, 399, 462
\bibitem[Bertin \& Arnouts 1996]{BA96} Bertin, E. \& Arnouts, S.
1996, \aap S, 117, 393
\bibitem[Blumenthal et al.\ 1984]{BFPR84} Blumenthal, G. R., Faber,
S. M., Primack, J. R., \& Rees, M. J.  1984, \nat, 311, 517
\bibitem[Bruzual \& Charlot 1993]{BC93} Bruzual, G. \& Charlot, S.
1993, \apj, 405, 538
\bibitem[Calzetti, Kinney, \& Storchi-Bergmann 1994]{CKS94}
Calzetti, D., Kinney, A. L., Storchi-Bergmann, T.  1994, \apj, 429, 582
\bibitem[Charlot \& Fall 1993]{CF93} Charlot, S. \& Fall, S. M.  1993,
\apj, 415, 580
\bibitem[Conti, Leitherer \& Vacca 1996]{CLV96} Conti, P. S., Leitherer,
C., \& Vacca, W. D.  1996, \apjl, 461, L87
\bibitem[Dressler \& Gunn 1992]{DG92} Dressler, A. \& Gunn, J. E.
1992, \apjs, 78, 1
\bibitem[Dressler et al.\ 1993]{DOBG93} Dressler, A., Oemler, A.,
Butcher, H., \& Gunn, J. E.  1993, \apjl, 404, L45
\bibitem[Dressler et al.\ 1994a]{DOBG94} Dressler, A., Oemler, A.,
Butcher, H., \& Gunn, J. E.  1994a, \apj, 430, 107
\bibitem[Dressler et al.\ 1994b]{DOSL94} Dressler, A., Oemler, A.,
Sparks, W. B., \& Lucas, R. A.  1994b, \apjl, 435, L23
\bibitem[Ebbels et al.\ 1996]{Ebbels96} Ebbels, T. M. D., Le Borgne, J.-F.,
Pell{\'o}, R., Ellis, R.S., Kneib, J.-P., Smail, I., \& Sanahuja, B.
1996, \mnras, 281, L75
\bibitem[Eggen, Lynden-Bell, \& Sandage 1962]{ELS62}  Eggen, O. J.,
Lynden-Bell, D., \& Sandage, A.  1962, \apj, 136, 748
\bibitem[Ellingson et al.\ 1996]{EYBE96} Ellingson, E., Yee,
H. K. C., Bechtold, J., \& Elston, R.  1996, \apjl, 466, L71
\bibitem[Fukugita, Shimasaku, \& Ichikawa 1995]{FSI95} Fukugita, M.,
Shimasaku, K., \& Ichikawa, T.  1995, \pasp, 107, 945
\bibitem[Giavalisco, Steidel, \& Macchetto 1996]{GSM96} Giavalisco, M.,
Steidel, C. C., \& Macchetto, D. F.  1996, \apj, 470, 189
\bibitem[Harris \& Racine 1979]{HR79} Harris, W. E. \& Racine, R.
1979, \araa, 17, 241
\bibitem[Holtzman et al.\ 1995]{Holtz95} Holtzman, J. A., Burrows,
C. J., Casertano, S., Hester, J. J., Watson, A. M., \& Worthey, G.
1995, \pasp, 107, 1065
\bibitem[Hu, McMahon, \& Egami 1996]{HME96} Hu, E. M., McMahon, R. G.,
\& Egami, E.  1996, \apjl, 459, L53
\bibitem[Hu \& McMahon 1996]{HM96} Hu, E. M. \& McMahon, R. G.
1996, \nat, 382, 231
\bibitem[Kelson et al.\ 1997]{Kelson97} Kelson, D. D., van Dokkum, P.,
Franx, M., Illingworth, G. D.  1997, in preparation
\bibitem[Kron 1980]{Kron80} Kron, R. G.  1980, \apjs, 43, 305
\bibitem[Leitherer, Robert, \& Heckman 1995]{LRH95} Leitherer, C.,
Robert, C., \& Heckman, T. M.  1995, \apjs, 99, 173
\bibitem[Lowenthal et al.\ 1997]{Lowenthal96} Lowenthal, J. D.,
Guzm{\'a}n, R., Gallego, J., Koo, D. C., Phillips, A. C., Vogt, N. P.,
Faber, S. M., Illingworth, G. D., Gronwall, C.  1997, \apj, in press
\bibitem[Madau 1995]{Madau95} Madau, P.  1995, \apj, 441, 18
\bibitem[Maeder \& Conti 1994]{MC94} Maeder, A. \& Conti, P. S.  1994,
\araa, 32, 227
\bibitem[Malkan, Teplitz \& McLean 1996]{MTM96} Malkan, M. A., Teplitz,
H., \& McLean, I. S.  1996, \apjl, 468, L9
\bibitem[McWilliam \& Rich 1994]{MR94} McWilliam, A. \& Rich, R. M.
1994, \apjs, 91, 749
\bibitem[Moore et al.\ 1996]{Moore96} Moore, B., Katz, N., Lake, G.,
Dressler, A., \& Oemler, A.  1996, \nat, 379, 613
\bibitem[Oke et al.\ 1995]{Oke95} Oke, J. B., et al.  1995, \pasp,
107, 375
\bibitem[Petitjean et al.\ 1996]{PPVC96} Petitjean, P., P\'econtal, E.,
Valls-Gabaud, D., \& Charlot, S.  1996, \nat, 380, 411
\bibitem[Pettini \& Lipman 1995]{PL95} Pettini, M. \& Lipman, K.
1995, \aap, 297, L63
\bibitem[Russell \& Bessell 1989]{RB89} Russell, S. C. \& Bessell,
M. S.  1989, \apjs, 70, 865
\bibitem[Sadler, Rich, \& Terndrup 1996]{SRT96} Sadler, E. M., Rich,
R. M., \& Terndrup, D. M.  1996, \aj, 112, 171
\bibitem[Searle \& Zinn 1978]{SZ78} Searle, L. \& Zinn, R.  1978,
\apj, 225, 357
\bibitem[Schuster et al.\ 1996]{SNPBO96} Schuster, W. J., Nissem,
P. E., Parrao, L., Beers, T. C., \& Overgaard, L. P.  1996, \aap
S, 117, 317
\bibitem[Seitz et al.\ 1996]{SKSS96} Seitz, C., Kneib, J.-P.,
Schneider, P. \& Seitz, S.  1996, \aap, 314, 707
\bibitem[Steidel et al.\ 1996a]{SGPDA96} Steidel, C. C., Giavalisco, M.,
Pettini, M., Dickinson, M., \& Adelberger, K. L.  1996a, \apjl, 462,
L17
\bibitem[Steidel et al.\ 1996b]{SGDA96} Steidel, C. C., Giavalisco, M.,
Dickinson, M., \& Adelberger, K. L.  1996b, \aj, 112, 352
\bibitem[Steinmetz \& Muller 1995]{SM95} Steinmetz, M. \& Muller E.
\mnras, 276, 549
\bibitem[Trager 1997]{Trager97} Trager, S. C.  1997, Ph.D. Thesis, UC
Santa Cruz
\bibitem[Vacca \& Conti 1992]{VC92} Vacca, W. D. \& Conti, P. S.
1992, \apj, 401, 543
\bibitem[Walborn et al.\ 1995]{Walborn95} Walborn, N. R., Lennon, D. J.,
Haser, S. M., Kudritzki, R.-P., \& Voels, S. A.  1995, \pasp, 107, 104
\bibitem[Yee et al.\ 1996]{Yee96} Yee, H. K. C., Ellingson, E.,
Bechtold, J., Carlberg, R. G., \& Cuillandre, J.-C.  1996, \aj, 111,
1783
\end{thebibliography}
\end{document}